\documentclass{cas-sc}
\usepackage{url}
\usepackage[numbers,sort&compress]{natbib}
\usepackage{framed}
\usepackage[most]{tcolorbox}
\usepackage{eurosym}
\usepackage{color}
\usepackage{float}
\usepackage{subfigure}
\usepackage{balance}
\usepackage{amsfonts}
\usepackage{paralist}
\usepackage{soul}
\usepackage{booktabs}
\usepackage{multirow}
\usepackage{graphicx}
\usepackage{lscape}
\usepackage{tikz}
\usepackage{multicol}
\usepackage{nomencl}
\usepackage{atbegshi}%
\usepackage{amsmath}
\usepackage{xcolor}

\DeclareDocumentCommand{\mytcolorbox}{O{} +m}{
    \begin{tcolorbox}[breakable, colback=white, height=0.5\textheight, #1]
            #2
    \end{tcolorbox}
}

\makenomenclature

\let\printorcid\relax 

\begin{document}
\def\floatpagepagefraction{1}
\def\textpagefraction{.001}
\shorttitle{Resilience assessment and planning in power distribution systems: Past and future considerations}

\title[mode = title]{Resilience assessment and planning in power distribution systems: Past and future considerations} 

\author[1]{Paul, S.}
\fnmark[$\dagger$]
\cormark[1]
\ead{shuva.paul@nrel.gov}

\author[2]{Poudyal, A.}
\fnmark[$\dagger$]
\cormark[1]
\ead{abodh.poudyal@wsu.edu}

\author[3]{Poudel, S.}

\author[2,3]{Dubey, A.}

\author[4]{Wang, Z.}

\address[1]{National Renewable Energy Laboratory, Golden, CO 80401, USA}
\address[2]{Washington State University, Pullman, WA 99164, USA}
\address[3]{Pacific Northwest National Laboratory, Richland, WA 99352, USA}
\address[4]{Iowa State University, Ames, IA 50011, USA}

\cortext[cor1]{Corresponding authors. National Renewable Energy Laboratory, Golden, CO, USA and  Washington State University, Pullman, WA, USA}
\fntext[fntext1]{Equal contribution authors.}

\begin{abstract}
Over the past decade, extreme weather events have significantly increased worldwide, leading to widespread power outages and blackouts. As these threats continue to challenge power distribution systems, the importance of mitigating the impacts of extreme weather events has become paramount. Consequently, resilience has become crucial for designing and operating power distribution systems. This work comprehensively explores the current landscape of resilience evaluation and metrics within the power distribution system domain, reviewing existing methods and identifying key attributes that define effective resilience metrics. The challenges encountered during the formulation, development, and calculation of these metrics are also addressed.
Additionally, this review acknowledges the intricate interdependencies between power distribution systems and critical infrastructures, including information and communication technology, transportation, water distribution, and natural gas networks. It is important to understand these interdependencies and their impact on power distribution system resilience. Moreover, this work provides an in-depth analysis of existing research on planning solutions to enhance distribution system resilience and support power distribution system operators and planners in developing effective mitigation strategies. These strategies are crucial for minimizing the adverse impacts of extreme weather events and fostering overall resilience within power distribution systems.
\end{abstract}


\maketitle

\nomenclature{HILP}{High-impact low-probability}
\nomenclature{ICT}{Information and communication technology}
\nomenclature{SA}{Situational awareness}
\nomenclature{DER}{Distributed energy resources}
\nomenclature{FOM}{Figure of merit}
\nomenclature{DG}{Distributed generators}
\nomenclature{MG}{Microgrid.}



\section*{Highlights}
\begin{itemize}
    \item Extreme event outages are primarily due to disruptions in power distribution systems
    \item There is no standard resilience metric; existing metrics vary in definition
    \item Resilience strategy integrates planning, operations, and recovery
    \item Interdependence of critical infrastructures is crucial for resilience planning
    \item Research gap exists in using distributed resources for grid resilience
\end{itemize}
\begin{multicols}{2}
\renewcommand{\nomname}{Abbreviations}
\printnomenclature
\end{multicols}

\section{Introduction}\label{sec:intro}
High-impact low-probability (HILP) extreme weather events, including hurricanes, heatwaves, wildfires, and more, are causing significant socio-economic impacts on the power grid~\cite{waseem2020electricity,xia2020research}. These events pose severe challenges to the operation and supply of the electric grid, affecting millions of customers~\cite{paul2019brief, 9281506}. For example, Hurricane Ian in 2022 affected approximately 2.7 million customers in Florida~\cite{nasa_observatory}, while Europe experienced an energy crisis during the 2022 heatwave, with France witnessing a drastic increase in electricity prices, reaching a record of \euro700/MWh~\cite{heatwave_europe}. The power outage in Texas in February 2021 left more than ten million people without power, causing a financial impact of about \$4 billion on wind farms, significantly more than their annual gross revenue~\cite{poulos2021ercot}. Iran experienced a summer heatwave in 2021 with temperatures exceeding 122 F, and a deficit of almost eleven gigawatts of electricity was reported~\cite{iran_news}. In 2021, Hurricane Ida caused widespread outages in the Northeastern US, affecting 1.2 million people, and it took almost fifteen days to restore the electric power entirely~\cite{ida}. In $2022$, there were eighteen different billion-dollar disasters in the United States alone, with an estimated loss of approximately $\$ 172 $ billion~\cite{ncei2021us}. Critical customers, such as hospitals and transportation systems, endure substantial economic losses during such calamities. Prolonged power outages also exacerbate vulnerabilities in personal safety and security, highlighting the need to increase resilience to extreme weather events.

\begin{figure}[t]
    \centering
    \subfigure[]
    {
        \includegraphics[width=0.48\textwidth]{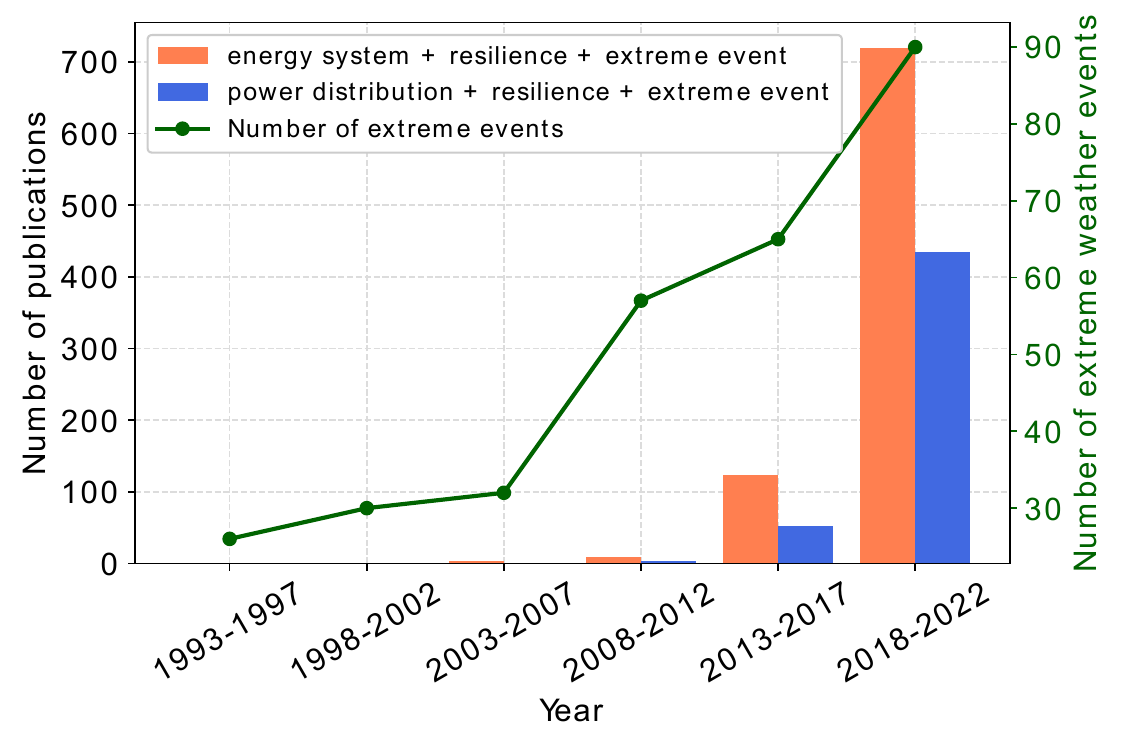}
    }\hfill
    \subfigure[]
    {
        \includegraphics[width=0.48\textwidth]{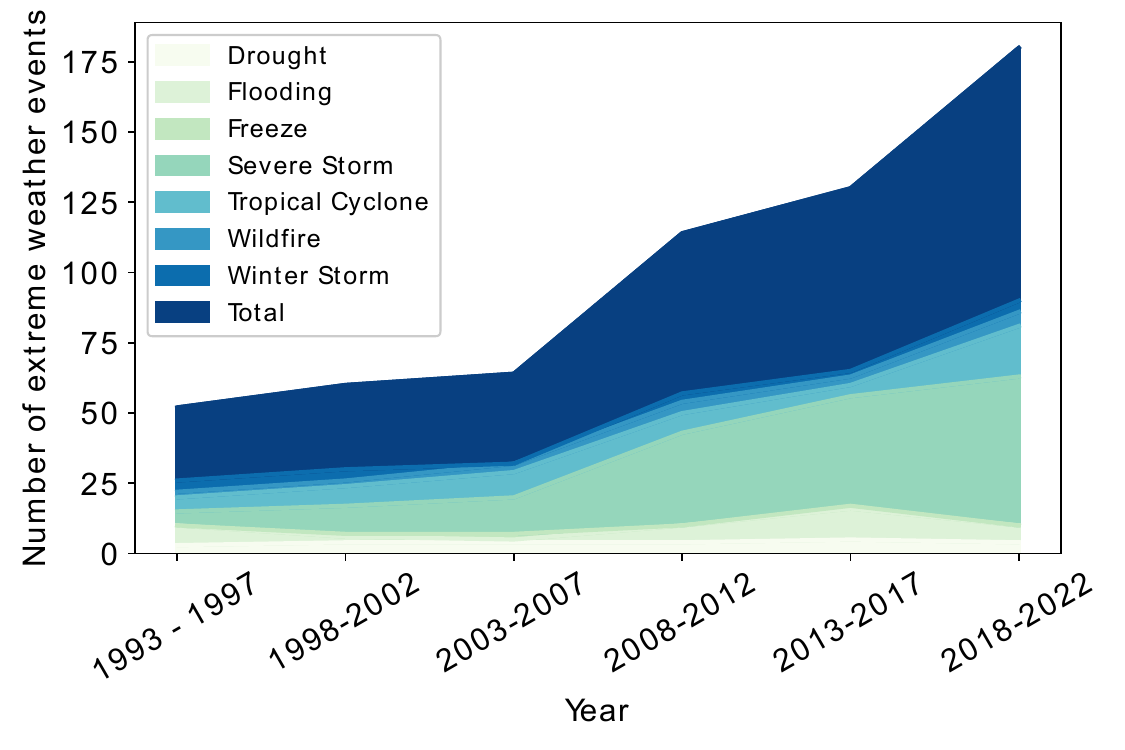}
    }\hfill
    \caption{a) Number of extreme events and trends in publications considering energy systems, power distribution systems, extreme events, and resilience since 1993 contained from the Scopus database. b) Different extreme events and their frequency every five years in the United States since 1993.}
    \label{fig:publication_events}
\end{figure}

The recent challenges faced by the power grid have highlighted the critical need for resilience due to extreme weather events. A resilient system can withstand severe disturbances, recover quickly to its normal operating state, and ensure uninterrupted power supply. It is worth noting that power distribution grids account for more than 80\% of power outages due to disruptions caused by extreme weather events~\cite{hou2023resilience}. Furthermore, due to the grid modernization initiatives, the bidirectional energy flow architecture of power grids, and the increasing impact of extreme weather events, ensuring the resilience of power distribution systems has become a pressing priority~\cite{GAO2021}. Although several questions and challenges remain, significant research efforts have been devoted to understanding and improving the resilience of power distribution grids. This review aims to comprehensively summarize existing research contributions in the field of resilient power distribution systems and identify key areas for future research, particularly in light of the anticipated increase in the frequency and severity of natural disasters. Figure~\ref{fig:publication_events}~a) shows the increasing number of research publications since 1993 on the impact of extreme events on energy systems, specifically related to power distribution systems. Figure~\ref{fig:publication_events}~b) illustrates the frequency of various extreme events in the United States. However, it should be noted that the increase in the frequency of extreme events is not limited to the US alone and is a global concern. The rise in publications reflects the urgency and importance of addressing the resilience of energy systems to extreme weather events. In particular, research on power distribution systems gained prominence in recent years, with advances in controllable distribution systems. Therefore, a comprehensive understanding of current state-of-the-art resilience assessment and planning methods in power distribution systems is crucial to realize practical implications, including enhanced operational resilience, effective risk management strategies, and informed infrastructure planning. This study also plays an important role in forming the development of regulations, standards, and guidelines related to resilience planning, emergency response, and strategic infrastructure investments. This allows policymakers, industry stakeholders, and researchers to address the challenges posed by extreme weather events effectively.

\begin{table}[t]
\centering
\caption{Comprehensive resilience study in the existing literature in different domains, their contributions, and limitations to alleviate the significance of this resilience study.}
\label{tab:my-table}
\begin{tabular}{@{}p{0.5in}p{1.2in}p{2.1in}p{2.1in}@{}}
\toprule
References & Domain & Summary & Drawbacks \\ \midrule
\cite{LABAKA201621} & \multirow{8}{*}{\begin{tabular}[c]{@{}l@{}}Critical infrastructure \\ systems\end{tabular}} & Holistic frameworks for developing critical infrastructure resilience & Missing distribution system resilience discussion \\ 
\cite{LIU2020106617,systems6020021} &  & Resilience assessment of multi-domain infrastructure systems & Very little analysis on power distribution system resilience, missing discussion of interdependencies \\
\cite{dick2019deep} & & Discusses the deployment of deep learning techniques in critical infrastructure resilience & Missing thorough and focused analysis of distribution system resilience, characterization, and interdependencies  \\ 
\cite{Nipa2019ComparativeAO,petrenj2013information,pursiainen2014towards,yu2011review,gay2013resilience,QUITANA2020101575,murdock2018assessment} & & Discusses critical infrastructure resilience including transportation, geoscience, and water distribution system & None of them have focused on electric power distribution systems resilience and their in-depth analysis \\ \midrule
\cite{doi:10.1063/1.5066264} & \multirow{10}{*}{Bulk power systems} & Discusses resilience enhancement strategies using microgrids & Missing focuses of distribution system resilience, resilience metric characterization, understanding the interdependencies of the infrastructures  \\
\cite{7893706} &  & Analyze a load restoration framework to enhance the resilience of the power system against an extreme event & Lacks metrics analysis, characterizing resilience, and other resilience enhancement strategies \\ 
\cite{7105972, CAI2018844} & & Comprehensive studies on power system resilience against natural disasters focusing on forecast models, corrective actions, and restoration strategies &  Missing characterization of resilience metrics \\ 
\cite{7409923} & & Discusses the resilience and security of smart grid infrastructures & Missing the interdependence and metric analysis  \\ \midrule
\cite{8586495} & \multirow{15}{*}{\begin{tabular}[c]{@{}l@{}}Power distribution \\ systems\end{tabular}} & Comprehensively summarizes the distribution system resilience assessment frameworks and metrics. &  Lacks characterization of resilience and resilience enhancement methods \\
\cite{willis2015measuring} &  & Discusses the definition, frameworks, resilience frameworks development, etc. & Missing the discussion of critical understanding of interdependencies for multi-domain resilience assessment and enhancement \\ 
\cite{8566253} & & Discusses resilience enhancement utilizing microgrids in distribution systems and definitions of resilience & Lacks understanding of the resilience characterization in distribution systems, other strategies of resilience enhancements, and understanding of the interdependencies \\
\cite{9062002, MISHRA2021110201} & & Comprehensive study of resilience assessment on the power distribution systems. & Missing resilience characterization, discussion on interdependencies of critical infrastructures, resilience analysis process, and enhancement strategies \\ \bottomrule 
\end{tabular}%
\end{table}

Table \ref{tab:my-table} summarizes some of the existing review studies on resilience-related topics in different critical infrastructure systems, including power distribution systems, and outlines their contributions and limitations. These survey articles include related work on resilience quantification and planning applied to different domains. A holistic framework for the development of critical infrastructure resilience interrupted by external and unexpected forces is discussed in~\cite{LABAKA201621}. The described framework builds the foundation for resilience on a set of resilience policies, considering the influences of policies on the damage prevention, absorption, and recovery stages, and presents implementation methodologies. Similarly, in~\cite{LIU2020106617}, a review of the resilience of six critical infrastructures is presented, including electric, water, gas, transportation, drainage, and communication networks. The resilience of critical infrastructure elements and their main factors are studied in~\cite{systems6020021} involving electricity, gas, information and communication technology (ICT), and road transportation networks. The deployment of deep learning techniques for critical infrastructure resilience is presented in~\cite{dick2019deep}.
Similarly,~\cite{Nipa2019ComparativeAO,petrenj2013information,pursiainen2014towards,yu2011review,gay2013resilience, QUITANA2020101575, murdock2018assessment} presents a comprehensive review of different critical infrastructure resilience, including, but not limited to, transportation, geology, and water-distribution system. However, none of the existing works specifically addresses resilience enhancement or assessment methods for electric power distribution systems, including the associated infrastructural and operational considerations. 

In the domain of power systems, existing articles extensively cover the resilience of the high-voltage bulk power grid~\cite{doi:10.1063/1.5066264, 7893706, 7105972, CAI2018844}. However, there is a noticeable gap in discussing resilience considerations for medium-voltage and low-voltage power distribution systems. Bulk power grids and power distribution systems drastically differ in structure and operation. Thus, the general understanding of the bulk grid system does not translate directly to distribution systems that require specialized analysis. Although some review works touch on the resilience of smart grid infrastructures~\cite{7409923}, resilience metric and quantification~\cite{willis2015measuring, 8586495}, microgrid-based resilience assessment and enhancement~\cite{8566253}, resilience planning against extreme weather events~\cite{shi2022enhancing}, and resilience assessment framework in power distribution systems~\cite{kandaperumal2020resilience}, there is a lack of comprehensive work which compiles the major aspects of resilience analysis, quantification, mitigation, enhancement, and multidomain interdependencies in power distribution systems.

Existing reviews on the resilience of the power distribution system lack several key aspects. First, there is a lack of a systematic framework for evaluating resilience. The existing works do not provide a comprehensive process for resilience analysis, which is one of the most important aspects for characterizing the resilience of any critical infrastructure, including power distribution systems. Furthermore, these works do not highlight resilience metrics specific to power distribution systems. Secondly, these review articles do not provide insight into the interdependencies among critical infrastructures that can significantly impact the resilience of the power distribution grid. The impact of extreme weather events on one critical infrastructure can have a drastic effect on the other, which can negatively impact the community. Finally, the research gaps and limitations discussed in existing works are broad, making it challenging to identify specific research directions. This study aims to address these gaps and comprehensively review all aspects of resilience in power distribution systems. This review provides valuable information to the scientific and engineering community in addressing the resilience challenges posed by extreme weather events. It is important to clarify that this study is focused solely on the resilience of the power distribution system and does not address other aspects of the resilience of the power system. By narrowing the scope, this work aims to provide a comprehensive analysis and practical recommendations tailored to enhance the resilience of power distribution systems. Specifically, the major contributions of this work are as follows.


\begin{figure}[!t!]
    \centering
    \includegraphics[trim ={0.7cm 0.7cm 0.7cm 0.7cm} ,clip, width=0.8\linewidth]{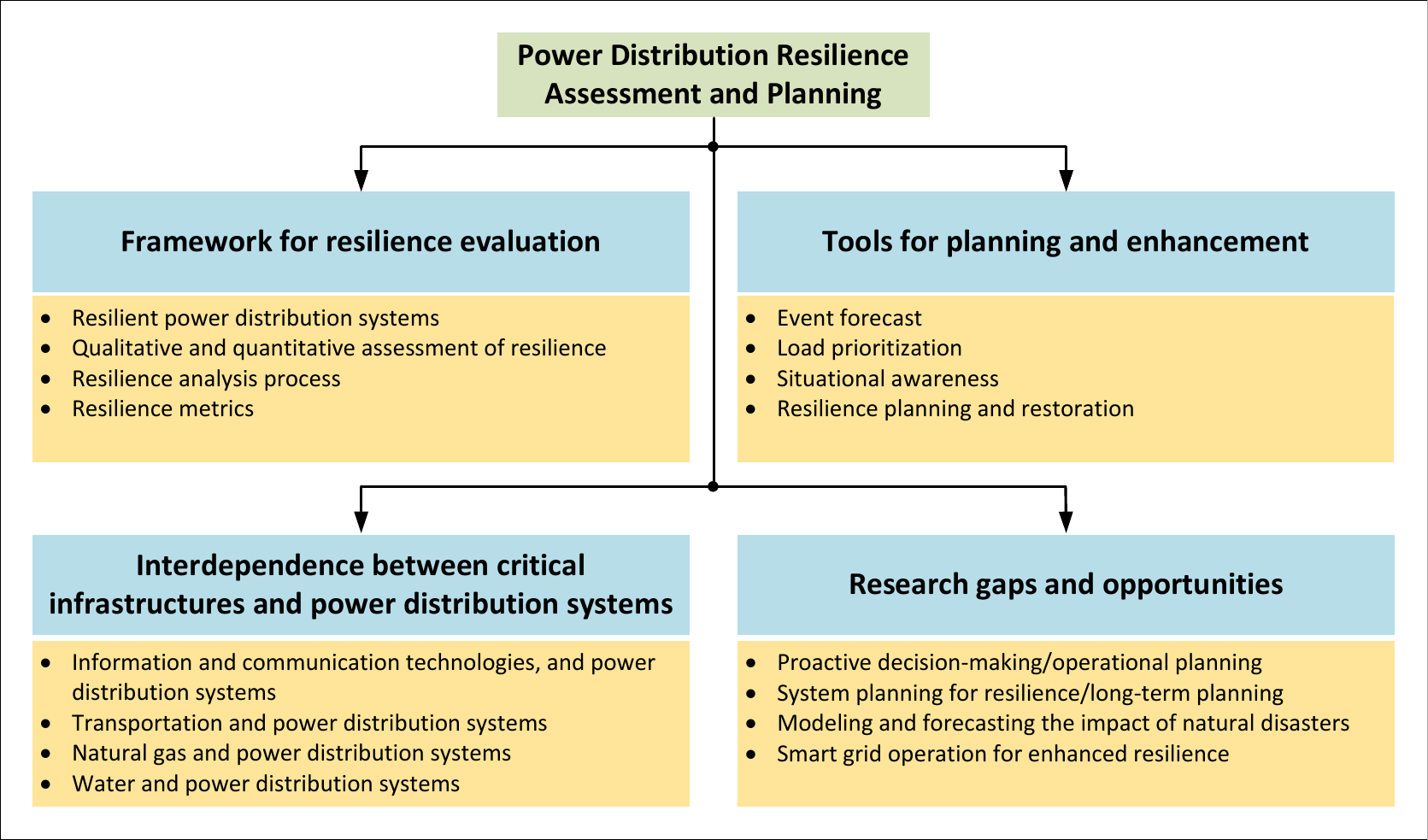}
    \caption{The overview of resilience assessment and planning in power distribution systems.}
    \label{fig:graphical_abstract}
\end{figure}

\begin{enumerate}
    \item The conceptual necessity for the resilience of the power distribution system is detailed, specifically in anticipation of HILP events, while also discussing the significance of existing definitions and their relevance.
    \item A characteristic representation of distribution system resilience is provided. This includes classifying assessments into qualitative and quantitative evaluations and attribute and performance-based metrics. Furthermore, a systematic resilience analysis process is detailed for power distribution systems. 
    \item A comprehensive survey of existing strategies to enhance the resilience of power distribution systems is presented. These strategies are categorized into event forecasting, load prioritization, situational awareness (SA), repair and resource allocation, and utilization of microgrids and distributed energy resources (DER) for resilience enhancement.
    \item The interdependencies among power distribution systems and various critical infrastructure systems, such as ICT, transportation, natural gas, and water distribution systems, are reviewed for their impacts on the resilience of the power distribution system.
    \item Research gaps and potential opportunities for further exploration in power distribution system resilience are identified and reviewed. These focus areas encompass proactive outage management, long-term resilience investment frameworks, investments in smart grid infrastructure, and modeling and forecasting the impact of extreme weather events.
\end{enumerate}

Figure~\ref{fig:graphical_abstract} shows the overview of the resilience assessment and planning in power distribution systems, and the rest of the work is organized as follows. Section~\ref{sec:res_eval} discusses methods to characterize power distribution grid resilience and resilience assessment techniques. Different types of resilience planning measures, standards, and operational procedures are discussed in section~\ref{sec:res_plan}. Section~\ref{sec:res_inter} focuses on the interdependence of power distribution systems with other critical infrastructures, followed by the potential research directions for resilience assessment in Section~\ref{sec:res_future}. Finally, Section~\ref{sec:conclusion} concludes the paper with a summary of the contributions of this study to the scientific community.

\section{Framework for resilience evaluation in power distribution systems}\label{sec:res_eval}
Resilience evaluation of power distribution systems is challenging due to the complex nature of distribution systems. Due to such intricacy, it is crucial to adopt a comprehensive approach that considers both qualitative and quantitative perspectives~\cite{kwasinski2016quantitative, kwasinski2012availability}. The resilience analysis process is vital in assessing the system's ability to maintain specific objectives during adverse grid conditions. Typically, the evaluation is quantified using some metric that aims to capture the system's resilience. However, it is widely acknowledged that a single metric cannot fully capture all the diverse resilience characteristics of power distribution systems~\cite{9708434}. Therefore, this section provides a comprehensive framework that explores various aspects of power distribution systems resilience, evaluation approaches, analysis process, and assessment metrics, aiming to facilitate a holistic understanding of resilience in power distribution systems.

\subsection{Resilient power distribution systems}\label{sec:res_rel}
\begin{figure}[t]
    \centering
    \includegraphics[width=0.9\linewidth]{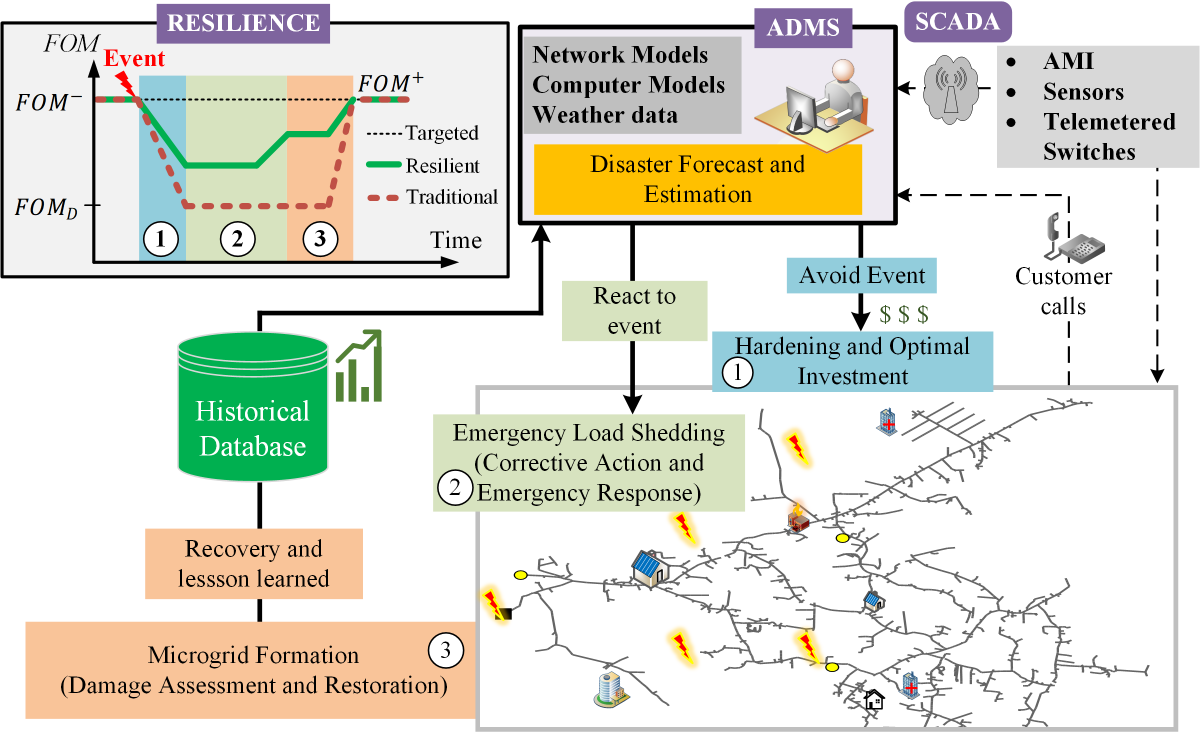}
    \caption{Response of power distribution system during HILP events.}
    \label{fig:resilience cycle}
    \vspace{-0.6cm}
\end{figure}

The response of a power distribution system when impacted by an extreme weather event is detailed in this section along with how it affects grid resilience. Figure~\ref{fig:resilience cycle} illustrates an overall response of the power distribution system, including some corrective actions after an HILP event. The top left portion of Figure~\ref{fig:resilience cycle} illustrates a general curve to show the response of a system towards an event. The vertical axis is the figure of merit (FOM) that accounts for the overall resilience of the system, and the horizontal axis represents time. FOM can be the number of customers online, the number of connected components, the electrical load fed by the system, etc. FOM is observed in every instance -- before, during, and after an extreme event and is commonly referred to as the system's resilience in this work. From Figure~\ref{fig:resilience cycle}, it can be seen that the system's resilience drastically decreases immediately after the event occurs. The appropriate infrastructure planning and hardening measures can slow down the rate of decrements in the system's resilience. When the system approaches the emergency response stage, a system with planned infrastructure shows better resilience than the traditional one with limited or no infrastructural development. An adequately hardened system with the knowledge of previous events that can effectively counteract the unwanted changes and restore the disrupted system in the lowest possible time is assumed to have higher resilience than the system that lacks these capabilities. Here, for the traditional system, $FOM^-$ represents FOM before the event, $FOM_D$ represents FOM before any remedial actions are taken, and $FOM^+$ represents FOM after corrective actions are taken.

To strengthen the effective response of the system to an event and minimize the potential impacts and necessary investments, it is essential to have an efficient long-term plan. One such investment is grid hardening~\cite{henderson2017electric}, which refers to making physical changes in its structure, e.g., replacing overhead lines with underground lines, elevating substations to prevent them from being flooded, and so forth. 
Although the effective response of the system might improve with planning measures, the occurrence, impact, and location of events are still not certain. Therefore, immediate corrective actions or emergency responses are needed during or after an event. Corrective action can include dispatch of mobile energy resources~\cite{yao2019rolling, xu2019resilience}, load shedding~\cite{nourollah2018coordinated, hosseinnezhad2018optimal}, and intentional islanding~\cite{8307765, 7857787, 7513408, 7438896, osti_1334056}. Intentional islanding can effectively provide continuous supply to critical loads in the system with the help of distributed generators (DG) while isolated from the main grid~\cite{7811272,https://doi.org/10.1002/2050-7038.12610, 9163237, radhakrishnan2020learning}. While critical loads are islanded, the operators can dispatch other generating units to pick up additional system loads. 
With improved impact modeling, the appropriate system response can be proposed to counteract the negative effect of the event. As seen in Figure~\ref{fig:resilience cycle}, all the impact modeling, long-term planning, and emergency response processes are interconnected and work together to improve the resilience of the distribution system. Advanced distribution management systems (ADMS) facilitate the interconnection of these processes by continuously interacting with the distribution grid through supervisory control and data acquisition (SCADA)~\cite{8949564}. 
Furthermore, the system operators are directly connected to the customers, who can provide additional information about the current situation of the distribution grid when an extreme event hits the grid.

\subsection{Qualitative and quantitative assessment of resilience} 

Qualitative assessment of resilience considers aspects of energy infrastructure, information systems, fuel supply chain, business structure, etc. The capabilities of the system involved in the qualitative resilience assessment include preparedness, mitigation, response, and recovery. Qualitative resilience assessment frameworks can guide policymakers in implementing long-term energy policies~\cite{7893706}. Existing work includes numerous qualitative resilience assessment methods. For example, the work in~\cite{carlson2012resilience,mcmanus2007resilience} assesses the resilience of the system and the regional level using questionnaires, investigations, and individual ratings to address individual, industrial, federal, and infrastructural resilience. A scoring matrix is developed in~\cite{ROEGE2014249} to evaluate the functionality of the system from different perspectives. The analytical hierarchy process is employed in~\cite{ORENCIO201362} to convert subject opinions into comparable quantities, eventually aiding operational decision-making.  

Quantitative assessment methods aim to numerically assess the resilience of critical infrastructure systems such as the power grid~\cite{JUFRI20191049}. Specific to power distribution systems, many studies have been conducted to assess system resilience quantitatively. The intrinsic characteristics of resilience can be defined as stress resistance and strain compensation. Later, the stress resistance is split into hardness and asset health. In contrast, strain compensation is characterized by capacity and efficacy~\cite{taft2017electric}. 
Another way to measure the resilience in the electric power grid that quantifies the efficacy of the recovery process is the ratio of recovered functionality to the system~\cite{6899257}. The resilience analysis process can be interpreted in multiple steps depending on time. The indices include expected hazard frequency, initial failure scale, maximum level of impact, recovery time, and recovery cost according to stages. A functional description of resilience is obtained in terms of initial failure scales, maximum impact level, and recovery time~\cite{doi:10.1061/41171(401)174}. 
When considering extreme weather events, the resilience assessment metric can be expressed as a function of the expected number of power grid outages during the event, the probability of loss of load, the expected demand not served, and the level of difficulty of the grid recovery processes~\cite{7489002}. 
Similarly, other resilience measures emphasize multiple system properties, including recovery time, loads not served, etc.,~\cite{doi:10.1193/1.1623497,PANT201492,doi:10.1111/j.0022-4146.2005.00365.x,NAN201735,doi:10.1111/risa.12093}.

\subsection{Resilience analysis process}
\begin{figure}[!ht!]
    \centering
    \includegraphics[width=0.85\textwidth]{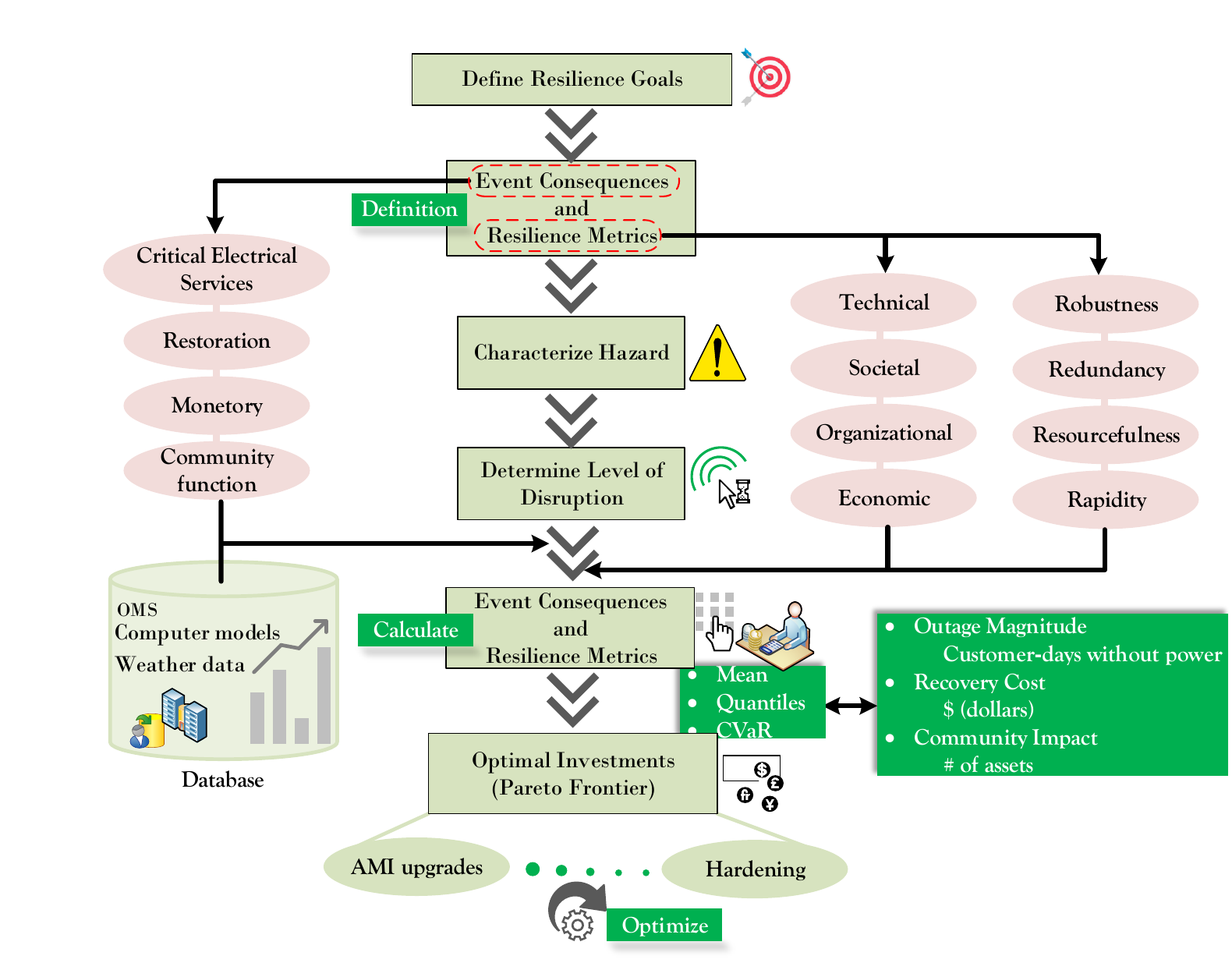}
    \caption{The resilience analysis process -- Setting up resilience goals, measuring resilience, and planning for resilience enhancement.}
    \label{fig:res}
\end{figure}

This section briefly discusses and extends the resilience analysis process, initially introduced for the 2015 quadrennial energy review in~\cite{watson2014conceptual}. The analysis framework evaluates the power system's capacity to handle potential future disturbances. It helps to prioritize planning decisions, investment endeavors, and response actions based on this assessment. In doing so, this study also highlights the available research and techniques that concentrate on every aspect of the analysis procedure to define resilience goals for utilities, choose suitable metrics that align with those objectives, collect essential data for the metrics, and ultimately determine the optimal approach to making resilience-based decisions. The conceptual framework and analysis process for creating forward-looking resilience metrics, which are based on extensive simulations that measure the impact on grid operations and power delivery, are shown in Figure~\ref{fig:res}. It provides a clear roadmap that outlines the journey from establishing resilience goals to effectively achieving them, with multiple interconnected components in between. 

\subsubsection{Resilience goals/objectives} 
The resilience analysis process starts with defining resilience goals. The goal could be to improve the resistance and capability of a regional electric grid to withstand extreme events, evaluate a utility's investment plan for resilience enhancement, reduce the cost of recovery (monetary and time), ensure power availability to critical services, and/or reduce the overall interruption cost. For example, in~\cite{kwasinski2016quantitative}, resilience is defined as the sum of the availability of individual critical loads during an event. In~\cite{ma2016resilience, ma2018resilience, 7381672}, the focus is on optimal investment decision-making capability to enhance resilience. Other studies have suggested various indicators of resilience, such as the optimal duration for repairing critical components~\cite{wen2020resilience}, disruptions in the energy supply following extreme events~\cite{espinoza2017seismic}, the general delivery of essential resources and uninterrupted power supply to critical customers after a disaster~\cite{poudel2018critical}, and the recovery of infrastructure and operations~\cite{gao2016resilience, umunnakwe2021quantitative}. Table \ref{tab:my_label} summarizes these indicators, and a comprehensive list of examples is available in~\cite{gmlc}. These indicators can be instrumental in defining resilience goals and developing relevant metrics to measure them.

\subsubsection{Hazard characterization}
The grid disturbance event could be initiated by various natural threats. Several efforts have been made to create better ways to evaluate the impacts of these threats. It is worth noting that different threats to energy infrastructure vary in probability and impact. For instance, a wind storm can cause significant damage to overhead structures but may not impact underground systems such as cables and substations. Consequently, it is critical to understand the hazard characterization as part of the resilience analysis process. Hazard US (HAZUS) or CLIMate ADAptation (CLIMADA) tool can be used effectively to model or simulate extreme weather events, including but not limited to hurricanes, floods, and earthquakes~\cite{vickery2006hazus, scawthorn2006hazus, Climada}. When evaluating resilience against multiple hazards effectively, it is crucial to consider two key factors: (1) the probability of each potential threat scenario and (2) how the intensity of an event maps onto the resulting consequences at the system level.

\subsubsection{Event to impact mapping}
Once the onset of the event is identified, the next step is to use the characteristics or attributes of the event to determine the impact at the component level. Different components in the power grid have predefined thresholds (intrinsic property of a device) to handle the subjected disturbances. These thresholds depend on the type of component and the duration of exposure. Although the thresholds are not directly measurable, historical data can be used to derive the failure probabilities~\cite{online0002}. These component-level failure probabilities can then be mapped to characterize system-level impacts. For example, a Monte Carlo simulation approach can be used to assess the spatiotemporal impacts of grid disturbances using fragility curves~\cite{8848458, panteli2016power, poudyal2021spatiotemporal}. Grid disturbance theory and functional form modeling capture the spatiotemporal effects of any possible scenarios~\cite{hanif2020modeling}. Although functional models can describe grid disturbance, event onset must be carefully analyzed to demonstrate the most meaningful impact and responses. This is crucial because different disruptive events can have varying effects on a system, requiring specific strategies for resistance and recovery. The work in~\cite{wang2022systematic} reviews some widely used impact models in energy systems planning and operations due to extreme weather events.

\subsubsection{Calculate consequence-- performance measure and resilience metrics}
In this stage of the resilience analysis process, the consequence category is defined, which forms the basis for the development of the metrics. Resilience metrics are then determined for each category of consequences related to the technical, societal, organizational, or economic impacts of an event. Some commonly used metrics are demand/energy not served, recovery time and cost, load recovery factor, revenue loss, and customers not served~\cite{afzal2020state, raoufi2020power}. Other indices, such as restoration efficiency, vulnerability index, degradation index, and microgrid resilience index, are also widely studied~\cite{amirioun2019metrics}. 
It should be noted that the impact of potential threats is dependent on the system's capacity to (1) anticipate, prevent, and mitigate them before being affected by the event, (2) adapt to, absorb, and survive the threat when it occurs, and (3) recover, restore, reconfigure, and repair itself afterwards~\cite{kandaperumal2020resilience}. Since a single resilience metric cannot capture all possible aspects of the threat response, these metrics are a function of resilience goals, the operating conditions and intrinsic characteristics of the system, and new investment planning initiatives.

\begin{table}[t]
    \centering
    \caption{Some widely used examples of the category of consequences and resilience metrics~\cite{vugrin2017resilience}.}
    \begin{tabular}{l l}
   \Xhline{3\arrayrulewidth}
        \\[-0.75em] 
       Category of consequence & Resilience metrics \\
        \\[-0.75em] 
       \Xhline{3\arrayrulewidth}
       \\[-0.75em] 
        Electric service & Total customer-hours of outage\\
         & Total customer energy demand not served\\
         & Average/percentage of customers experiencing a power outage during a specified period\\
        \\[-0.75em] 
        \hline
        \\[-0.75em] 
        Recovery & time and cost of recovery\\
        \hline
        \\[-0.75em] 
        Monetary & Loss of revenue, cost of damages, repair and resource allocation\\
        & Loss of assets and business interruption cost\\
        & Impact on gross municipal/regional product\\
        \\[-0.75em] 
        \hline
        \\[-0.75em] 
        Community function & Critical services without power\\
        & Critical customer energy not served\\
        \\[-0.75em] 
       \Xhline{3\arrayrulewidth}
    \end{tabular}
    \label{tab:my_label}
\end{table}

\subsubsection{Evaluate resilience improvements}
Once the resilience metrics are established, the potential investment for resilience enhancement is studied. Given limited resources and high risk, resilience can be improved by focusing more on low-probability events, investment prioritization, and consequence management. These studies allow system analysts to decide on investments based on evolving research to identify the most impactful decision in improving resilience while minimizing long-term costs and stranded investment~\cite{schwartz2019utility}.
In recent years, it has been realized that investments should focus on quantitative analysis, the ability to incorporate the uncertainty of grid disturbance, and bottom-up approaches where efforts for resilience enhancement should start from the grid-edge~\cite{taskrepo01}.

\subsection{Resilience metrics}
This section details the desired characteristics of a metric defining distribution system resilience and the categories of resilience metrics for distribution system resilience planning. A metric is essential for resilience planning, as it quantifies the impacts of potential HILP events on the grid and helps evaluate and compare planning alternatives to improve operational resilience~\cite{mukhopadhyay2016public, national2017enhancing, shandiz2020resilience}. Measuring progress towards a more resilient infrastructure requires developing and deploying metrics that can be used to assess infrastructure planning, operations, and policy changes~\cite{petit2013resilience,carlson2012resilience}. The resilience metric should focus on HILP events, consider the likelihood and consequences of threats, and evaluate the system's performance. Furthermore, the metric should consider the uncertainties inherent in response and planning activities while quantifying the consequences of power grid failures~\cite{watson2014conceptual, force2022methods}.

It is incredibly challenging to adopt a unified metric that can capture several contributing factors such as uncertainty, spatiotemporal features of a threat, and intrinsic system properties to deal with possible threats~\cite{ton2015more, aki2017demand}. Current resilience measures can be classified into two main types: a) attribute-based metrics, which assess the attributes of the power system such as adaptiveness, resourcefulness, robustness, SA, and recoverability~\cite{kandaperumal2020resilience}, and b) performance-based metrics, which evaluate the ability of the system to remain energized (commonly referred to as availability~\cite{cai2018availability}), often represented by the conceptual resilience curve~\cite{panteli2017power}.

\subsubsection{Attribute-based metrics}

Attribute-based metrics are relatively simple in mathematical formulation, and the required data collection is also easier than performance-based metrics. The fundamental question that attribute-based metrics aim to answer is ``What makes the system more/less resilient than other systems?". Attribute-based metrics are used to provide a baseline understanding of the system's current resilience and are driven by the properties that increase the resilience of the concerned system. 
The properties of the system comprise robustness, resourcefulness, adaptivity, recoverability, and SA. For example, the ratio of underground feeders to overhead feeders, the proportion of distributed resources to critical consumers, the number of advanced metering infrastructures/sensors, path redundancy, and overlapping branches result in increased robustness, resourcefulness, and SA, thus improving the resilience of the system to HILP events~\cite{kwasinski2016quantitative, bajpai2016novel}. Some of the widely used attribute-based resilience metrics are described below:
\begin{enumerate}
    \item System robustness: This metric evaluates the ability of the power distribution system to withstand shocks or disturbances without failure. For example, the robustness of the system can be evaluated based on the strength of the system infrastructure, such as the resilience of poles, wires, and transformers to severe weather events~\cite{panteli2016power}.
    \item System flexibility: This metric evaluates the system's ability to adapt to changes or disturbances. For example, the flexibility of the system can be evaluated based on its ability to manage power supply and demand during peak hours, integrate renewable energy sources, and/or extract demand flexibility during scarcity~\cite{hanif2023analyzing}.
    \item System redundancy: This metric evaluates the system's ability to maintain power supply even when one or more system components fail. For example, redundancy can be evaluated based on the number of backup power sources, such as generators or batteries, available to maintain the power supply during outages~\cite{dong2018battery} or tie switches to utilize the power from the feeder~\cite{poudel2018critical}.
    \item Customer satisfaction: This metric evaluates the ability of the power distribution system to meet customer expectations during and after disruptions. For example, customer satisfaction can be evaluated based on the system's ability to provide timely and accurate information during outages and the quality of customer service provided by the system's operators.
\end{enumerate}

The definition of resilience plays an important role in the evaluation of resilience metrics based on the attributes of the system~\cite{watson2014conceptual}. When forming the attribute-based resilience metric, the spatiotemporal features of an HILP event on power distribution systems are also taken into account~\cite{7091066,7893706}. The uncertainty associated with HILP events is a critical attribute that is often represented using probabilistic measures~\cite{7046691}. These attributes, incorporated into the metric, are valuable for decision-making in planning and policy-making processes~\cite{watson2014conceptual, 7893706,ton2015more}. When utilizing attribute-based metrics, it becomes possible to compare different systems, both with and without resilience enhancement strategies. Attribute-based resilience enhancement metrics need to be able to adapt to advances in technology, ensuring their continued relevance and effectiveness. 

\subsubsection{Performance-based metrics}
It is recommended that the grid resilience metrics are defined based on system performance. They should be (1) forward-looking, (2) quantify the consequences of disruptions, (3) incorporate uncertainties that can affect the response of the system and planning decisions, and 4) be flexible enough to use data from historical analysis and system models~\cite{vugrin2017resilience}. Such performance-based metrics follow the approaches in evaluating system resilience quantitatively. The performance of the electric grid during major shocks, such as natural disasters, can be described by outage frequency, the number of customers impacted, outage duration, or a combination of these. The national energy regulatory commission (NERC) published separate metrics to evaluate system performance against reliability standards~\cite{lauby2012nerc, NERC_standard}. The growing occurrence of extreme events has emphasized the importance of developing metrics to assess the performance of the power system during HILP events. Eventually, entities are expected to appropriately include all events that affect the power system, considering the event's probability and impact on the communities. 
Some of the widely used performance-based resilience metrics are described below:
\begin{enumerate}
    \item Energy at risk: Energy at risk is a metric that quantifies the amount of energy that may not be supplied during extreme events. It provides a forward-looking assessment of the potential consequences of disruptions by estimating the amount of energy that can be lost due to outages during extreme events \cite{gruber2022profitability}.
    \item Probabilistic assessment risk: Probabilistic assessment risk is a metric that assesses the likelihood and consequences of disruptions due to various failure scenarios. It reflects inherent uncertainties by considering the probability of various failure scenarios and the potential consequences of these failures \cite{8848458}. It can use historical analysis and system modeling to quantify the likelihood of different scenarios and the potential consequences of these events.
    \item Flexibility margin: This metric measures the ability of the system to respond to changes in demand and supply. It is forward-looking and reflective of inherent uncertainties by assessing the system's ability to respond to unexpected changes in demand and supply, especially during scarcity and emergency conditions \cite{roege2014metrics}.
    \item Restoration time: Restoration time is a metric that measures the time it takes for the power system to restore power following a disruption \cite{hosseini2020quantifying}. It is forward-looking and quantifies the consequences of disruptions by assessing the time required to restore service to customers. Historical analysis and system modeling can be used to estimate the time required to restore customer service in various scenarios.
\end{enumerate}

A resilience performance curve is widely used to define performance-based metrics~\cite{POULIN2021107926}. Several studies have adopted a resilience triangle in the past to determine the system performance where only two different states are presented~\cite{MISHRA2021110201}. Lately, it has been realized that the one-dimensional character of the resilience triangle is not very helpful and can only capture the recovery from an event. It is equally important to capture other highly critical resilience dimensions such as ``how fast resilience degrades and how long the system remains in the degraded state before the recovery stage"~\cite{7091066, nichelle2021extracting}. A conceptual resilience curve has recently been used to assess and define performance-based metrics~\cite{fang2016resilience, panteli2017metrics}.

In conclusion, performance-based approaches are commonly used in cost-benefit analysis and planning studies to assess the advantages and drawbacks of proposed resilience improvements and investments. While attribute-based and performance-based approaches have distinct definitions and can be used independently depending on utility preferences, combining these approaches allows for a more comprehensive analysis of grid resilience, considering the potential consequences of disturbances. Attribute-based metrics provide a broad characterization of grid resilience, while performance-based approaches assess tailored options for enhancing resilience, integrating economic, social, and regional factors. By combining attribute-based and performance-based metrics, a baseline evaluation of resilience, recovery efforts, planning, and investment activities can be effectively maximized, leading to improved grid resilience~\cite{poudyal2022metric, 9708434}. Unlike these methods, some recent works also explore a data-driven method to characterize resilience in power distribution systems~\cite{zhu2021quantifying}. Data-driven approaches are likely to be popular as critical information from the electric utility is less likely to be publicly available.    

\begin{table}[t]
\caption{State-of-the-art-- Resilience planning and enhancement strategies for power distribution systems.}
\centering
\label{tab:tab3}
\begin{tabular}{@{}ccc@{}}
\toprule
References & Resilience category & Planning \& enhancement strategy \\ \midrule
\cite{7765016} & \multirow{6}{*}{Operational resilience} & Load prioritizing \\
\cite{6457435, lubkeman2000field, atanackovic2013deployment,gray2015making, jia2013state, huang2015evaluation, 7562822} &  & Situational awareness \\
\cite{8758803,li2014distribution, gao2016resilience, wang2015self,chen2017modernizing, chen2015resilient, wang2016three, farzin2016enhancing,chen2017sequential, chen2017multi, article} & & Microgrids and DERs-based load restoration \\ 
\cite{nguyen2019assessing} & & Mobile energy resources \\ 
\cite{8355692, 9115714, 9265243, 8010861} & & Deployment of soft open point technology \\ \midrule
\cite{SGIGR,alliance2014future, sandy, PNNL} & \multirow{3}{*}{Infrastructural resilience} & Remote units deployment \\
\cite{lei2016mobile, xu2019resilience, lei2018routing} &  & Deployment of mobile energy resources \\
 &  &  \\ \midrule
\cite{xu2015placement} & \multirow{3}{*}{\begin{tabular}[c]{@{}c@{}}Both operational \& \\ infrastructural resilience\end{tabular}} &  Smart distribution systems \\
\cite{8640043,8587147,8587140} &  & Repair scheduling, Optimal switching, crew dispatching \\
~\cite{wang2015networked, arif2017networked} &  & Networked microgrid \\  
~\cite{arif2017power} & & Repair and restoration using DGs, switches, and crews\\ \midrule  \bottomrule 
 
\end{tabular}%
\end{table}

\section{Power distribution system resilience -- Tools for planning and enhancement}\label{sec:res_plan}
\begin{figure}[!t!]
    \centering
    \includegraphics[width=0.5\textwidth,trim={0.5cm 0.5cm 0.5cm 0.5cm}, clip]{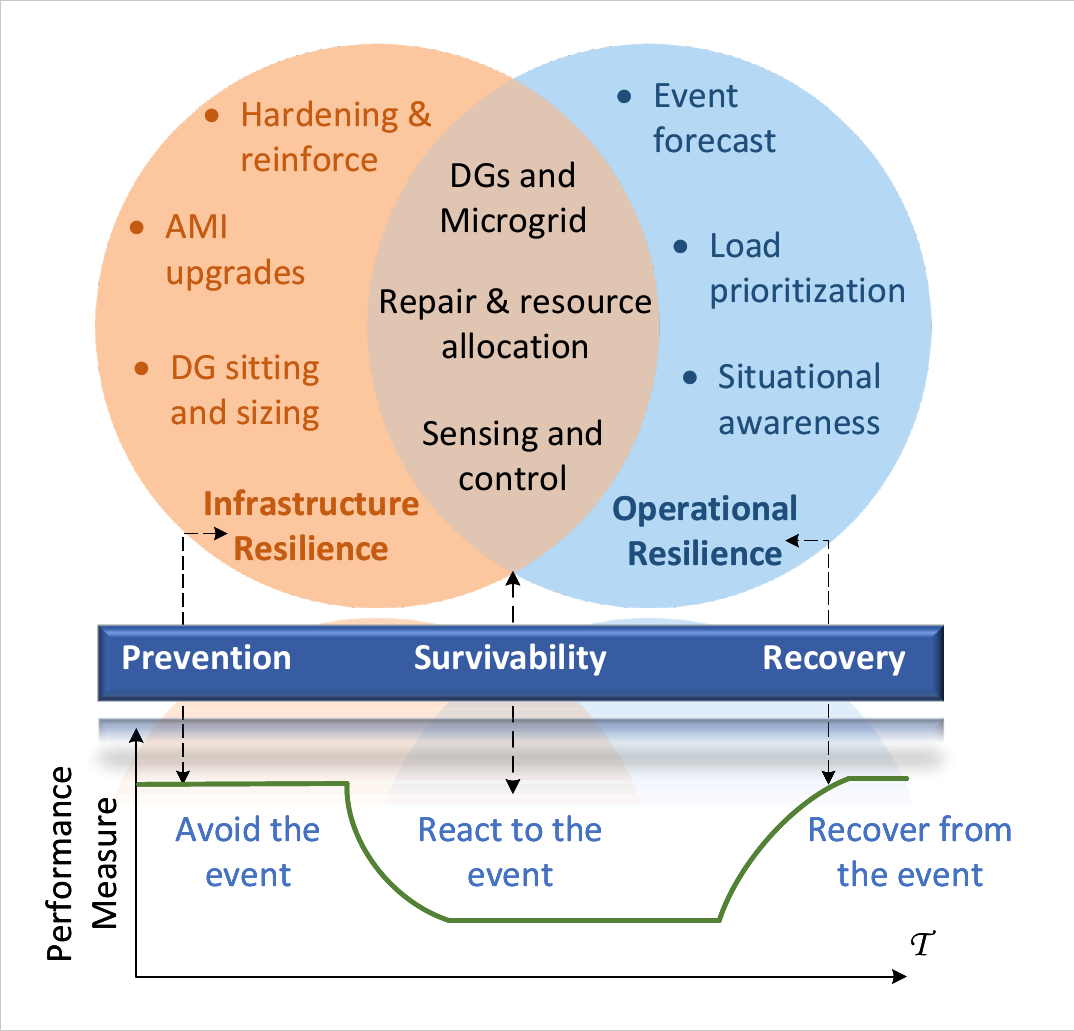}
    \caption{Different stages of resilience planning and enhancement strategies.}
    \label{fig:tools}
\end{figure}

The resilience enhancement of power distribution systems depends on various tools and strategies. These include accurately forecasting natural events, recognizing critical loads that require uninterrupted power supply, maintaining situational awareness, and implementing proper planning and restoration measures. The combination of these measures and tools contributes to the overall enhancement of system resilience. These measures can be further categorized into operational and infrastructural resilience, as detailed in Table~\ref{tab:tab3}. The stages of a resilience curve --- avoid, react, and recover --- are illustrated in Figure~\ref{fig:tools}, indicating where these tools can be applied to enhance resilience~\cite{hanif2020modeling}. This section details related work on critical measures and tools for resilience planning and enhancement in power distribution systems. 

\subsection{Event forecast}
Extreme weather forecasting helps utility planners make appropriate operational decisions to reduce damage to the power grid. Recently, advances in observation networks such as satellite remote sensing have significantly improved the accuracy of short-term weather forecasting models~\cite{shanmugapriya2019applications}. Additionally, advances in data analytics, accurate weather modeling, and enhanced computing resources make extreme weather forecasting efficient and reliable~\cite{NOAA_forecast}. However, long-horizon weather prediction is still an active area of research and requires further consideration when used for infrastructural planning purposes. Nevertheless, with advancements in short-term predictions, utility planners can take advantage of accurate event forecasting to make appropriate operational decisions. These decisions may include resource scheduling, dispatching crews and other resources for maintenance, and stocking backup resources. Essentially, a weather-grid impact model can be developed to understand and simulate the effects of an extreme weather event on the power grid. The existing research on the weather-grid impact model can be categorized into statistical and simulation-based models. These models include a detailed modeling of power systems, extreme weather events, damage assessment, and restoration after extreme natural events~\cite{cerrai2019predicting,4349087,guikema_sandy}. 

\subsection{Load prioritization}
Hazardous incidents that impact the operation of the distribution system disrupt the supply to critical loads, thus affecting grid resilience. To ensure a high level of resilience for critical infrastructures, federal and state authorities are actively working to identify and provide guidance for enhancing critical infrastructure security~\cite{cis}. Critical customers such as hospitals, fire departments, water suppliers, and other emergency units are recognized and prioritized for the prompt restoration of power supply within the utility's service area. This underscores the significance of enhancing the resilience of critical infrastructures through the increased availability of essential customers and their critical loads. The objective is threefold: 1) facilitate a rapid response to grid disturbances, 2) reduce the magnitude of harm and challenges experienced by communities, and 3) accelerate the restoration of critical functions~\cite{chalishazar2021power}.
Moreover, due to the significant amount of grid outages caused by catastrophic events, it is impossible to prevent threats to all assets. Therefore, prioritizing the critical assets of the grid becomes a clear choice for utilities~\cite{gmlc}. Critical loads identification allows operators to selectively disconnect low-priority loads and maintain backup resources within their generation capacity while sustaining vital facilities for an extended duration~\cite{hampson2013combined, marnay2015japan}. In such cases, various advanced techniques can be employed to restore the prioritized loads while considering topology and operational constraints, thereby enhancing the overall system resilience~\cite{DING2017205}.

\subsection{Situational awareness}
It is crucial to have adequate SA of the system conditions for effective and timely decision-making to counteract the impacts of the HILP event~\cite{6523855}. Several incidents discussed in Section~\ref{sec:intro} illustrate the severity when the system operators are not fully aware of relevant information. Inadequate SA increases the probability that the system enters the cascading blackout phase~\cite{6457435}. In light of these events, several industrial stakeholders have significantly advanced information systems to enhance SA. One key element of this is state estimation, which is critical to enabling continuous and reliable monitoring and control of the distribution systems, particularly in the presence of DER penetration~\cite{qin2023integrated}. To date, the observability of the system downstream of the substations is very limited. Hence, only a limited number of utility companies have implemented control room applications, including, but not limited to, state estimation, topology identification, fault management, and cyber attack mitigation~\cite{atanackovic2013deployment}. Furthermore, it is unclear how to adapt existing SA technologies from the bulk power system to power distribution due to unknown network models, poor observability, and incorrect measurements~\cite{bajagain2022iterative}.

Nevertheless, progress in power distribution systems has brought together various innovative technologies like digital relays, phasor measurement units, intelligent electronic devices, automated feeder switches, voltage regulators, and smart inverters for DER. These advancements pave the way for improved end-to-end awareness and visualization of the network. Furthermore, the regular polling and on-demand retrieval of customer interval demand data through the advanced metering infrastructure contribute to enhancing the accuracy of the online model for the distribution network~\cite{jia2013state, huang2015evaluation}.
The integration of advanced metering infrastructure information in SA tools is supplemented with conventional supervisory control and data acquisition, which provides additional data to improve the system's SA~\cite{boardman2010role, 7095588}.
In the future, most of the electric power grid will have installed communication networks, intelligent monitoring, and distributed sensors along feeders to provide additional data for an improved SA~\cite{epri}. It is also envisioned that such digital networks will ultimately lead to greater levels of communication between the end-users, the utilities, and with other physical infrastructures~\cite{campbell2018smart}. The use of drones and unmanned air vehicles for damage assessments is also gaining popularity~\cite{7778146}. The improved SA and increased automation in a smart grid paradigm will assist the control room operators with real-time decision-making, thus improving the grid resilience~\cite{stevens2015situation}. 

\subsection{Resilience planning and restoration}
Resilience planning and restoration involve various strategies and actions aimed at improving the ability of a distribution system to withstand and recover from extreme events. These strategies can be broadly categorized into three main areas: long-term resilience investment, short-term pre-event preparation, and post-event restoration. 

\subsubsection{Long-term resilience investment}
Long-term resilience investment involves strategic planning to make the distribution system more resilient to uncertain and extreme events. This includes infrastructure reinforcements or system hardening methods such as installing underground lines, elevating substations, and other upgrades to improve the system's reliability and robustness~\cite{Alguacil2014, Yuan2014, 5948404}. However, unlike transmission systems, distribution systems have unique characteristics, such as radial topology, low redundancy, and inability to incorporate DC power flow methods, which require specific considerations in resilience planning~\cite{7381672}. Distribution systems have received less attention compared to transmission systems, with limited literature on resilient distribution system design~\cite{8241850, Hughes2021, 9209142}. 

Most of the studies in resilience planning and upgrades apply two different types of modeling techniques, namely robust~\cite{7381672} and stochastic modeling~\cite{7021963,yamangil2015resilient,ma2019resilience, gao2017resilience}. The scenario-based stochastic methods and other network interdiction models facilitate optimal hardening strategies in the distribution system~\cite{7381672, ma2018resilience, ma2016resilience, Meera2021}. In other works, DGs siting/sizing and automatic switch placement strategies are simultaneously formulated to minimize the overall expected cost~\cite{9121323,9006872}. However, with the growing need of resilience enhancement, investment decisions should be based on HILP events rather than the expected cost of several possible events. Furthermore, resource planning should be carried out to fulfill the need for operational flexibility, and specialized power distribution system models should be integrated with advanced operations~\cite{watson2014conceptual, Watson2010}. 
Works have shown that sensing and control technologies can also be deployed in the planning phase to enhance the resilience of the distribution system~\cite{8082533,8787584}. Integrating these resources in the planning phase and observing the trade-off between these resources against uncertain or extreme events can provide a better planning portfolio.

\subsubsection{Short-term pre-event preparation}
Short-term pre-event preparation focuses on resource allocation and planning strategies that can be implemented just before an extreme weather event to enhance the resilience of the distribution system. This includes activities such as pre-staging of resources, crew dispatching, and network reconfiguration to minimize the impact of the event and expedite restoration. One approach to short-term pre-event preparation is the strategic placement of resources, such as emergency response generators and crews, in strategic locations before an extreme weather event occurs. This allows for quicker deployment and utilization of resources immediately after the event, reducing the time required for restoration. For example, the U.S. Federal Emergency Management Agency (FEMA) pre-stages emergency response generators before hurricanes or other major events, but the effectiveness of this strategy depends on the accuracy of the event forecast and the optimal placement of resources~\cite{fema}. Mathematical models, such as mixed-integer linear programming models, can be used to optimize the pre-staging of resources based on various factors, such as forecasted event severity, expected outage duration, and available resources~\cite{8640043,8587147,8587140}.

Short-term preparation also involves proactive network reconfiguration, where the distribution system is strategically reconfigured so that the damaged sections can be quickly isolated and power can be restored to unaffected areas once the event occurs. This can be achieved through automated switching actions or remote-controlled switches that can be operated remotely, allowing for quicker restoration without needing physical presence on site~\cite{lei2017remote, xu2015placement}. Optimization models can be used to determine the optimal switching actions and reconfiguration plans considering system topology, available resources, and outage duration~\cite{8758803, 8640043}. Furthermore, repair crew deployment and coordination is an essential preparation plan to ensure efficient restoration efforts. Crews must be strategically deployed to affected areas based on event severity, expected restoration time, and available resources. Mathematical models can be used to optimize crew deployment and scheduling to minimize restoration time and maximize the utilization of available resources~\cite{arif2017power}. There are other efforts in pre-allocating mobile resources as a short-term preparation process~\cite{9372331, lei2016mobile, lei2018routing}. The main idea is to restore critical loads by forming small MGs with the available mobile resources for a resilient emergency response to natural disasters. Such intentionally islanded MGs will have voltage support grid-forming resources that can energize the islanded loads until the restoration and repair tasks are completed. 

\subsubsection{Post-event restoration}
Post-event restoration involves activities carried out after an extreme weather event to restore the distribution system to normal operation. It is one of the most critical components of a system's resilience, as quick and effective restoration justifies an efficient planning strategy. Hence, planning and pre-event preparation are vital in ensuring a quick and efficient restoration process. Restoration includes repairing damaged infrastructure, re-configuring the network, and optimizing allocated resources to expedite restoration~\cite{poudel2018critical}. The repair crews must assess the damage, repair or replace damaged components, and restore the system to its normal operating condition following an event to minimize downtime and expedite restoration. Network reconfiguration also plays a crucial role in post-event restoration. The distribution system must be reconfigured to restore power to affected areas and isolate damaged sections. This can be achieved through automated switching actions or remote-controlled switches. Mixed integer programming models can be leveraged to optimize the repair and restoration process by coordinating crew assignments, resource allocation, and network reconfiguration to minimize restoration time and ensure efficient utilization of available resources~\cite{8640043,8587147,8587140}. 

The research on post-event restoration focuses also on the resilience enhancement against natural disasters using DERs and MGs, with varying restoration objectives~\cite{li2014distribution, wang2015self}.
MGs provide an effective solution on DERs management and utilization for system restoration after extreme weather events~\cite{ schneider2020slider, osti_1576926, 8913212, 8973466, 8618618, 8606281, 8440136}. Networked MGs consider dispatchable as well as non-dispatchable resources for service restoration in power distribution during long-duration outages~\cite{wang2015networked, arif2017networked}. MGs can also be adaptive in which the formation of MG and load switching sequence is guided by the nature of extreme events~\cite{che2018adaptive}. MGs and DERs can also be used for sequential service restoration~\cite{chen2017sequential, chen2017multi}. The DG dispatch and network switching can be coordinated well to generate a feasible restoration sequence. Furthermore, such restoration can also be performed in multiple or hierarchical stages~\cite{arif2017power,farzin2016enhancing}. Service restoration can be achieved via dynamic changes in the boundaries of MGs within a distribution network that includes synchronous-machine DGs~\cite{kim2016framework,chen2017modernizing}. It is vital to maintain the grid frequency throughout the restoration process~\cite{xu2017dgs}. The correlation between switching actions and frequency deviations is considered and a suitable switching sequence is formulated that meets the essential requirement of adhering to the dynamic frequency nadir limit. Some controllers, such as grid-friendly appliance controllers, can avoid large transients in low inertia microgrids associated with switching operations for coordination~\cite{7857787}.

\section{Interdependence between critical infrastructures and power distribution systems}\label{sec:res_inter}
\begin{figure}[t]
    \centering
    \includegraphics[width=0.88\textwidth,trim={0.88cm 1.05cm 0.05cm 1.85cm}, clip] {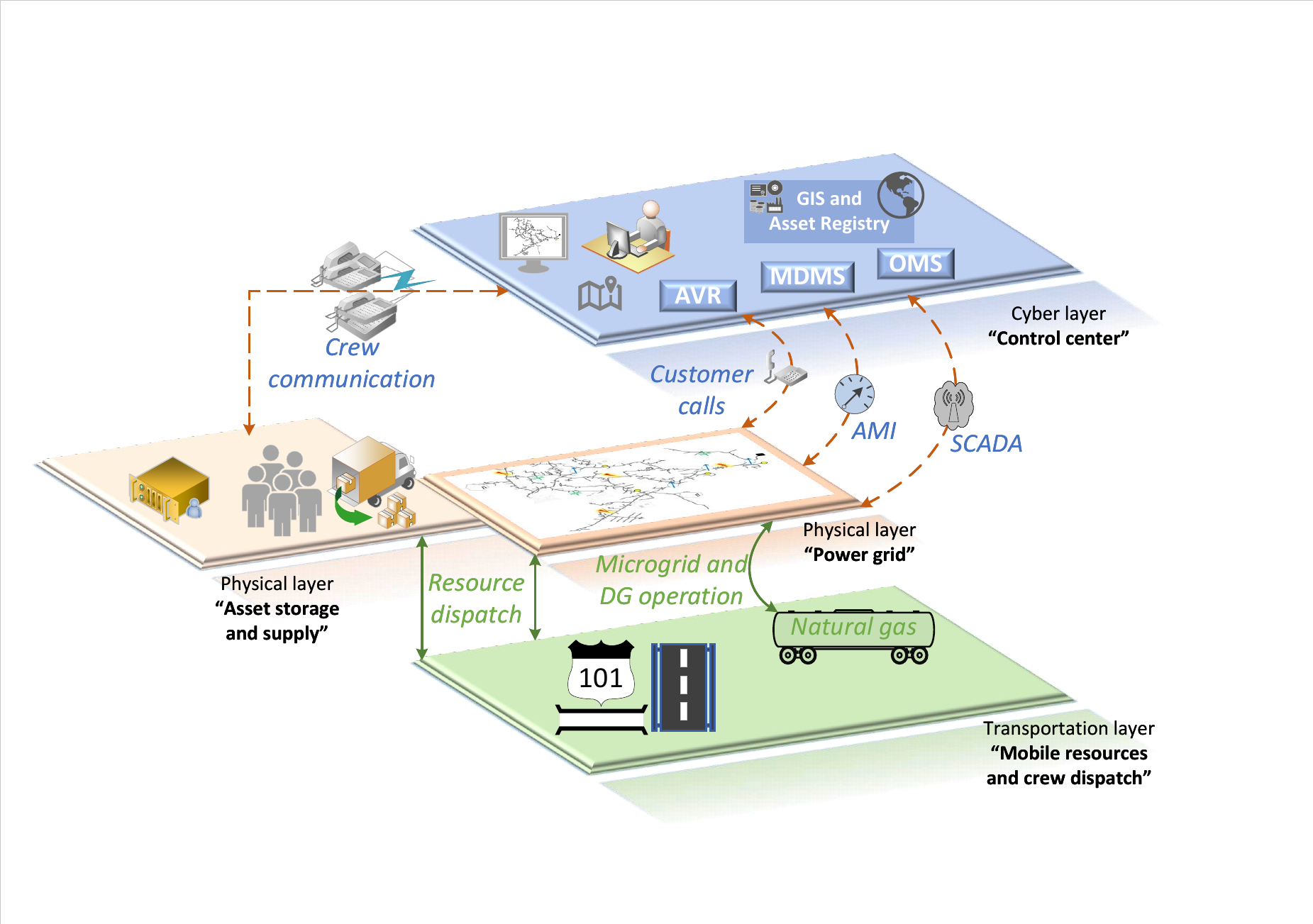}
    \vspace{-0.5 cm}
    \caption{Interdependence among various critical infrastructures, namely ICT network, transportation network, natural gas network, and physical power grid.
    }
    \label{fig:layered}
    \vspace{-0.3 cm}
\end{figure}

The inherent interdependencies between power distribution systems and other critical infrastructures contribute to the resilience of the community~\cite{en12101874}. Figure~\ref{fig:layered} shows a high-level overview of the interdependencies of critical infrastructures with the power distribution system. It is crucial to understand these interdependencies for effective disaster response and recovery planning, as disruptions in the power distribution system can have cascading effects on other critical infrastructures, amplifying the overall impact on the community~\cite{fema_cascade}. However, there is little understanding of the complex dynamics, vulnerabilities, and emerging threats associated with these interdependencies. This section investigates and analyzes these infrastructure interdependencies, highlights the contributions that have been made so far in this problem space, and addresses some open challenges.

\subsection{Information and communication technologies, and power distribution systems}
Interdependent power distribution systems and ICT networks offer opportunities to mitigate vulnerabilities and leverage infrastructural convergence. Key trends in analyzing their interdependencies include increasing data volumes, faster system dynamics, hidden feedback, renewable portfolio standards, variable energy resources penetration, network cybersecurity, co-simulation needs, reliability coordination, and real-time/data-in-motion analytics. Figure~\ref{fig:layered} illustrates the interdependencies between cyber layers and other physical layers of the power distribution system. Prioritizing cybersecurity is crucial for federal research and industry. Investigating these interdependencies helps to develop assessment tools to specify ICT requirements for advanced grid functionalities and build a strong foundation for new grid management tools. The deployment of advanced sensing and measurement technologies facilitates data collection and understanding while improving the resilience of the system with strategic decisions. The transmission of data streams, such as measurement data, from field devices to the control center for monitoring, analysis, and control purposes faces security risks, including data leaks, hacks, and adversarial intrusion~\cite{SENGAN2021107211, Olowu2021}. There are four domains --- cyber, social, energy market, and distribution system networks --- that interact to enable advanced grid functionalities and revolutionize grid modernization. However, such interaction also exposes the network to severe security threats. 

Failure or hacking in the cyber layer can have cascading effects on the physical layer, affecting equipment and services. Understanding the relationship between the distribution grid's architecture and control systems is crucial. During extreme weather events, the power grid becomes highly vulnerable. If cyber intruders breach the control and communication system when impacted by an extreme event, it can critically collapse the power grid~\cite{8881730}. Loss of communication makes damage assessment and asset management nearly impossible, leading to incomplete situational awareness. Limited situational awareness during extreme events hinders decision-making. Additionally, considering cyber layer constraints can enhance the resilience of the power system, as the cyberinfrastructure is interconnected with the physical layer~\cite{8558579}. Advanced distribution management systems and networked microgrid control paradigms should coordinate to address operational conflicts and reconnect microgrids to the distribution networks after events. In addition, the future of distribution management systems will see an increased dependence on distributed resources and distributed architecture, necessitating robust communication capabilities~\cite{molzahn2017survey}. 

   
\subsection{Transportation and power distribution systems}
It is essential to model the critical infrastructure dependency between power distribution and transportation, as the dispatch of the crew and other mitigation processes can be delayed if the transportation network is not accessible when the region is affected by an extreme event. In Figure~\ref{fig:layered}, the interdependency between the transportation layer and the physical layer (power grid) is represented. The transportation system and the operational dependency of the power grid play a critical role in the resilience enhancement during extreme weather events, especially when preparing for an upcoming storm by allocating mobile resources (mobile energy storage, poles, distribution lines, etc.) and dispatching repair crews for optimal service restoration, calling for thorough research in understanding and modeling the related interdependencies.  During extreme weather events, critical operations, such as optimal resource planning for rapid system restoration and emergency evacuation mechanisms, require proper modeling of the transportation network associated with the power distribution system. To investigate the critical elements that need upgrading or expansion, the study of the influence of contingency on traffic flow and power flow can be a great addition to interdependence modeling~\cite{8982274}.
Additionally, the optimal selection of emergency stations (including distribution centers, power supply recovery, emergency supplies, medical centers, etc.) requires specific attention during interdependence modeling considering disaster management~\cite{PEREZGALARCE201756,BOONMEE2017485,RAWLS2010521}. Such an interdependence assessment can greatly help disaster-impacted areas develop a rapid recovery plan. For example, after Hurricane Maria in Puerto Rico, it took $11$ months to restore the power back to its nominal state. One of the primary reasons for such a lengthy restoration was the lack of proper modeling of the transportation systems, preventing efficient crew dispatch~\cite{hurricane_maria}.

\subsection{Natural gas and power distribution systems}

The primary reasons for the interdependence between natural gas pipelines and power distribution systems are attributed to the space heating using natural gas in residential, commercial, and industrial buildings. Simultaneous damage to both networks due to a natural disaster can cause severe concern to affected communities. Another example includes rare winter heating days, when gas distribution utilities may enforce price spikes, leading to reliability challenges. For instance, in the $2004$ New England cold snap, the real-time price of electricity rose the ISO-New England's bid cap~\cite{england2004final}. The day-ahead gas prices at New England's gas system were increased (nearly ten times their normal price range) along with the electricity price. The interdependence should be modeled to jointly identify the overall patterns of electricity generation and distribution, focusing on the energy requirements of electricity and natural gas. The framework for interdependence between natural gas and power distribution systems requires assessment of integration and automation in these infrastructures (such as threat/hazard identification and data acquisition), identification of potential impact zones, initial and cascading effects on infrastructural assets from failure events, and identification of propagation paths of this disruptive events~\cite {doi:10.1061/(ASCE)IS.1943-555X.0000395}. Furthermore, power distribution systems energize the constituent components of the gas distribution system. The operation of natural gas processing plants and other relevant assets, including electricity-powered compressor stations, depends on the power distribution networks~\cite{4077098}. Failures in the corresponding power distribution network may propagate to the gas distribution network due to the strong interdependencies~\cite{CHERTKOV2015541}. 

\subsection{Water and power distribution systems}
The manageability and resilience of power and water distribution system are challenged by their increasing interdependence and inter-connectivity, as widely studied within water-energy nexus activities~\cite{doi:10.1061/(ASCE)WR.1943-5452.0001000}. While power system research has made considerable progress over the years through dedicated research efforts and active community participation, the tools used in existing studies on water distribution networks are comparatively less advanced~\cite{singh2018optimal}. Critical infrastructures such as water and power distribution systems are widely interdependent since they share energy, computing, and communication resources. In heavily loaded power distribution systems, failures on distant power lines can cause severe water supply shortages. The water distribution network comprises several hydraulic components, including pipes, pumps, and tanks. The power system energizes some of these elements. Any failure will trigger cascaded loss and interrupt the electrical power supply to the hydraulic components. Research is being carried out to understand and assess this interdependence using different performance metrics~\cite{9707888}. One such metric is the demand satisfaction ratio, which measures the impact of power failure on the interconnected water distribution network~\cite{10.1145/3397776.3397781}. The security framework of such critical interdependent infrastructures must be modeled using different state-of-the-art techniques, such as game-theoretic methods~\cite{8088670}. The operation of multi-purpose reservoirs in water distribution networks requires interdependent resource allocation between water and power distribution networks; it can be solved as a multi-objective optimization problem~\cite{GONZALEZ2020114794}. Furthermore, multi-infrastructural interdependence modeling paves the way for a more sophisticated resilience analysis for power distribution systems~\cite{en12101874}.


\section{Research gaps and opportunities}\label{sec:res_future}
There is a significant research gap in standardizing resilience quantification, modeling, and planning for critical infrastructures such as power distribution systems.
There are several areas where extensive research can improve the resilience quantification and analysis process. The research gaps specific to the analysis and enhancements of power distribution system resilience are summarized in the following sections.

\subsection{Proactive decision-making/operational planning}
From the utility's perspective, planning for an upcoming HILP event is desirable to enhance resilience proactively. Planning includes mechanisms to reduce the impacts of an upcoming event or resource allocation to assist in faster recovery. For instance, staging the repair crews at appropriate locations, analyzing the availability of supply resources, and deciding on an appropriate restoration scheme before the event can help reduce their impacts on the system and help accelerate recovery~\cite{arif2017power}. The necessity and effectiveness of proactive decision-making have been extensively discussed for both the bulk grid~\cite{7862805, 7725489, huang2017integration} and distribution systems~\cite{7539274, 8082545, gholami2017proactive,8839073, arif2017power}. In general, proactive outage management can help the system recover and restore faster in the aftermath of an event, thus reducing the overall impact of the HILP event. However, solving the resulting problem requires addressing some crucial modeling and algorithmic challenges. The uncertain and time-varying nature of HILP events must be appropriately modeled in the problem formulation. Robust algorithms are needed to solve related decision-making problems that typically involve solving stochastic non-linear (and often) mixed-integer optimization problems, which are computationally expensive. It is also important to model the spatiotemporal characteristics of the available resources (both human and automated) in the proactive decision-making process. For instance, fully utilizing a currently available resource for a specific outage can impact the operation in the future due to the inherent stochasticity of the resource and the changing nature of the event. In such a case, an optimal solution will include multi-stage planning to obtain optimal allocation at a given time step, considering future requirements and uncertainties. This is an even more complex problem to scale computationally for large-scale nonlinear systems such as the power grid. Additional research is needed to efficiently solve the resulting problem considering all the uncertainties related to the event and resources while appropriately modeling the complex operational decision-making problems of large-scale power systems.

\subsection{System planning for resilience/long-term planning}
Resilience planning is a long-term goal for the efficient and robust operation of electric power distribution systems. Investment plans are required for infrastructure hardening, including vegetation management, smart device installations, pole maintenance, upgrades, etc. Investments are required to install weather stations in high-risk areas to avoid high-impact extreme events. State-level regulators need to reevaluate their efforts in prioritizing investments for resilience enhancement by utility companies and reassess several techniques of resilience assessment from the perspective of regulatory decisions which might impact state-level grid investments (such as DERs). Although electric utilities are supposedly investing approximately $\$1$ trillion in the U.S. electric power grid between $2020$ and $2030$, investments must be implemented so that economic and national security perspectives can promote resilience by design. Utility companies require significant investment to improve security against potential vulnerabilities in distribution systems. These investments can make the distribution grid more resilient to HILP events~\cite{reimgrres}.

However, several uncertainties are associated with the planning process; inappropriately incorporating those can lead to sub-optimal investment decisions. Therefore, the planning problem should appropriately model such uncertainties and the associated risks imposed on the power delivery systems. Additionally, it is also important to model those risks in the optimization framework to achieve realistic outcomes from the planning process. Recently, some current work incorporates a convex risk measure, namely conditional value-at-risk, when solving the distribution system restoration problem~\cite{ Wu2018, Mahzarnia2020, 9094201}. Similarly, other works explore the uncertainties associated with the system in the development of restoration approaches~\cite{8750846, 8879675, arif2018optimizing}. However, the risk associated with the uncertainties should also be determined for a more accurate resilience assessment. Hence, while modeling the stochastic nature of events or available resources, it is equally important to model associated risks~\cite{munoz2017does}.

The long-term planning problem presents significant challenges in terms of modeling, solving, and scalability~\cite{homer2020electric}. Unlike operational planning, long-term planning requires the consideration of multiple scenarios and depends on the number of scenarios and time horizon being considered. Additionally, long-term planning decisions must account for the full profile of extreme weather events, necessitating a multi-hazard model that encompasses various events occurring independently or simultaneously~\cite{staid2021critical}. Furthermore, it is crucial to have a robust forecast model with minimal errors to justify future planning decisions~\cite{heydt2011characterization}. However, one major challenge in multi-hazard modeling and forecasting is the need for a large number of scenarios to represent such hazards. Balancing computational efficiency and accuracy is critical, as incorporating sufficient scenarios is necessary to capture the high uncertainty associated with HILP events. The long-term planning model typically includes stages before, during and after the event, but the complexity grows exponentially with the time horizon, number and nature of resources, and size of the power grid under consideration. Uncertainties arise from factors such as future weather conditions, solar irradiance, battery state-of-charge, and operating conditions of DERs. To address these challenges, sophisticated tools for scenario generation and reduction are required to model the problem while retaining essential information effectively. Future research should also focus on scalable algorithms and leveraging high-performance computing resources to enhance computational tractability~\cite{ISONE_HPC,zhao2014evaluation}.

\subsection{Modeling and forecasting the impact of natural disasters}
The recent and rapid changes in weather and the frequent occurrence of natural disasters are alarming, as these events can have long-lasting impacts. Therefore, modeling and forecasting the effects of extreme weather events play a major role in the resilient operation of the power distribution system. Most of the previous works consider hurricanes as extreme events while assessing the resilience of the distribution system. On the contrary, the different types of extreme weather events differ significantly in their impact on power distribution systems. Hence, considering all the parameters affecting the distribution system resilience, a generalized impact modeling framework is challenging. Additionally, there is a gap in appropriately modeling the impacts of extreme weather events on power distribution systems when designing system hardening solutions. Most impact models are based on the topology of the power system, lacking details about localized geographical information~\cite{7105972}. Weather forecasting is vital for routine operations, balancing production and demand. It is essential to warn of extreme events, making it possible to better manage demand and supply, prepare a response, and accelerate recovery times. Utilities lack improved models that downscale global information to the local level. Comprehensive research is required to develop tools and skills to interpret the data and understand how meteorological uncertainty affects current and future operations.
Moreover, since the frequency of these events is extremely low, the available data to characterize these events are limited. Novel methods are needed to use limited data to model the impacts at the component and system level of extreme weather events. There is also a critical need for accurate predictive event and damage models that use limited data. Weather data, meteorological data, historical outage reports, and other useful data sources must be integrated to improve resilience planning and preparation models for an upcoming event. Furthermore, the study should also have provisions for integrated studies with multiple critical infrastructures~\cite{8088670}. The interdependencies between power distribution and other critical infrastructures, as discussed in section~\ref{sec:res_inter}, will be crucial, as there is a lack of a comprehensive impact assessment model on interdependent infrastructures due to extreme events.   

\subsection{Smart grid operation for enhanced resilience}
Improved SA and controllability at the distribution level provide an additional venue to enhance resilience through smart grid operations. In recent years, advanced metering infrastructure and intelligent electronic devices such as sensors and telemetered controllers have been deployed in distribution systems, providing utilities with access to large and increasing amounts of data~\cite{9502840, 8166777}. Furthermore, phasor measurement units, which provide real-time synchrophasor data, are expected to be widely deployed in distribution grids~\cite{liu2014optimal}. These innovative grid-edge technologies enhance SA compared to traditional approaches, which are labor-intensive and time-consuming. However, it is not practical to successfully deploy such devices across the network due to the associated cost and geographical difficulties~\cite{tnd}. Therefore, a cost-benefit analysis is needed for the design and deployment of improved state estimation technologies. This can enhance system monitoring performance and facilitate model validation with post-event analysis, allowing for accurate decision-making~\cite{kezunovic2010new}. 

Another aspect of resilient smart grid operation is enabling non-traditional ways of operating grids using microgrids and other DER such as PV, storage, and flexible loads. 
Recently, DERs have been extensively examined for resilience enhancement because they support critical loads during extreme events independent of the bulk power system in an islanded mode. These DERs can also be effectively engaged with the help of transactive energy systems. Transactive energy systems can address operational challenges during abnormal conditions when new mechanisms are designed for contingencies \cite{bhattarai2021transactive}. Although conventional approaches for resilience enhancement are usually prepared from the system operator's standpoint, the transactive energy systems mechanism, if appropriately designed, can be utilized to incentivize customers to engage in activities that shift load to where it is needed the most and reduce the peak loads, thus relieving stress on the grid during scarcity \cite{dong2021integrating, huang2018simulation, reeve2022distribution}.

\section{Conclusions}\label{sec:conclusion}
Resilience planning and assessment for power distribution systems have emerged as promising but challenging research topics within the community. The intricacies associated with these systems require a thorough investigation of the resilience assessment, quantification, and analysis processes. This work aims to highlight the significance of examining the resilience analysis process of power distribution networks during extreme weather events. As this review evolved with the strategic discussion and analysis of the resilience assessment and analysis process and the relevant challenges associated with different domains, it became increasingly evident that ensuring a robust and resilient power grid is not only an expected characteristic but also an absolute requirement. It should be noted that certain aspects of power system resilience, such as transmission system resilience, were beyond the scope of this work. Future research should explore these areas to further advance the understanding and implementation of holistic power system resilience strategies. To summarize, the following key aspects are addressed in this review:
\begin{itemize}
    \item A comprehensive review is presented, highlighting the state-of-the-art resilience assessment processes and their limitations within the context of power distribution systems. This includes an overview of resilience assessment and quantification methods, an introduction to resilience analysis frameworks, and an examination of existing resilience metrics.
    \item The aspects of resilience planning in power distribution systems are thoroughly discussed. This encompasses event forecasting, load prioritization, SA, and resource planning and allocation. The interdependence of critical infrastructure systems is also analyzed, highlighting the interconnectedness between power distribution systems, ICT networks, transportation systems, natural gas distribution systems, and water distribution networks.
    \item Finally, critical research gaps are identified, and potential opportunities are proposed for future contributions in this field. By highlighting these gaps, it aims to guide current and future research to address the pressing challenges faced by the power systems community.
\end{itemize}

In conclusion, this research provides a comprehensive overview of resilience planning and assessment in power distribution systems, offering valuable insights and paving the way for further advancements in this critical area. The outcomes of this work have significant implications for multiple stakeholders in the energy sector. Industry stakeholders can leverage the resilience analysis framework to enhance their resilience planning, infrastructure investments, and operational strategies to mitigate the impacts of extreme events, as existing utility planning measures do not incorporate risks~\cite{jeremy_report}. Policymakers can utilize the insights to shape regulations, standards, and policies that promote the resilience of power distribution systems, aligning with global energy, environment, and sustainability goals. Furthermore, the research underscores the urgency of coordinated action at the international level to address the increasing frequency and severity of extreme events, emphasizing the need for collaborative efforts to build resilient energy systems that contribute to a sustainable and climate-resilient future. 

\section*{Declaration of Competing Interest}
The authors declare that they have no known competing financial
interests or personal relationships that could have appeared to influence the work reported in this paper.

\section*{Acknowledgement}
This work is supported by the U.S National Science Foundation (NSF) under grant ECCS-1944142.

\bibliographystyle{vancouver}
\bibliography{ref}

\medskip

\end{document}